# Worldwide scaling of waste generation in urban systems

Mingzhen Lu[1,2*], Chuanbin Zhou[3], Chenghao Wang[2], Robert B. Jackson[2,4], Christopher P. Kempes[1*]

**Author affiliations:**
[1]Santa Fe Institute, Santa Fe, NM 87501, USA
[2]Department of Earth System Science, Stanford University, Stanford, CA 94305, USA
[3]Research Center for Eco-Environmental Science s, Chinese Academy of Sciences, Beijing, 100085, China
[4]Woods Institute for the Environment and Precourt Institute for Energy, Stanford University, Stanford, CA 94305, USA

**Corresponding authors**:
Mingzhen Lu,
Santa Fe Institute,
1399 Hyde Park Rd, Santa Fe, NM 87501, USA
Email: mingzhen.lu@santafe.edu
ORCID https://orcid.org/0000-0002-8707-8745

Christopher P. Kempes,
Santa Fe Institute,
1399 Hyde Park Rd, Santa Fe, NM 87501, USA
Email: ckempes@gmail.com

**Summary paragraph**
The production of waste as a consequence of human activities is one of the most fundamental challenges facing our society and global ecological systems [1–4]. Waste generation is rapidly increasing[5], with corresponding shifts in the structure of our societies where almost all nations are moving from rural agrarian societies to urban and technological ones [6,7]. However, the connections between these radical societal shifts and waste generation have not yet been described. Here we apply scaling theory [8–11] to establish a new understanding of waste in urban systems. We identify universal scaling laws of waste generation across diverse urban systems worldwide for three forms of waste: wastewater, municipal solid waste, and greenhouse gasses. We show that wastewater generation scales superlinearly, municipal solid waste scales linearly, and greenhouse gasses scales sublinearly with city size. In specific cases production can be understood in terms of city size coupled with financial and natural resources. For example, wastewater generation can be understood in terms of the increased economic activity of larger cities, and the deviations around the scaling relationship - indicating relative *efficiency* - depend on GDP per person and local rainfall. We also show how the temporal evolution of these scaling relationships reveals a loss of economies of scale and the general increase in waste production, where sublinear scaling relationships become linear. Our findings suggest general mechanisms controlling waste generation across diverse cities and global urban systems. Our approach offers a systematic approach to uncover these underlying mechanisms that might be key to reducing waste and pursing a more sustainable future.

**Main text**

The production of waste as a fundamental aspect of living systems has characterized the history of the biosphere (including the oxygenation of the atmosphere as a photosynthetic byproduct[12]), inspired evolutionary transitions[13,14], and constrained the ecological dynamics of all temporal and spatial scales. The balance and cycling of several key greenhouse gasses such as carbon dioxide are largely defined as the waste products of biological and industrial metabolism[15–17]. For humans, the management of waste is a central consideration for health, well being, quality of life, impacts on the environment, efficient economies, and climate change[5,18–21], and these considerations have motivated everything from sewage systems to environmental regulations to the handling of medical and nuclear waste.

Human society is currently characterized by rapid population growth and urbanization, and thus, the ability to quantify and forecast the mechanisms behind urban waste production and reduce waste have tremendous benefits for policy and planning, strategic technological developments, and ecological modeling. However, the systematic connections between shifts in waste production and urbanization have not yet been described. The challenge is that frameworks for waste need to account for the nonlinear effects associated with city size in order to understand the fundamental mechanisms of waste production.

Scaling theory is an effective way to illuminate systematic behavior across diverse systems and to reveal novel mechanisms is scaling theory, specifically considering how features of a system change with its size [9]. Scaling theory has been successful in a variety of biological applications ranging from organism physiology to the structure of forests and mammalian ecosystems[8,9,22–24]. In many cases the scaling exponents between various features and size can be derived from fundamental physical, physiological, architectural, and structural limitations revealing fundamental mechanisms.

More recently, various efforts have applied scaling theory to quantify mechanistic processes operating in urban environments[9–11,25]. In urban systems, scaling has been used to describe everything from the production of new knowledge to shifts in inequality with city size[11,26,27]. Cities typically exhibit sublinear, linear, or superlinear scaling as characterized by the exponent *β* (Methods). Sublinear scaling (scaling exponents <1) often results from the efficiencies of scaling up (economy of scale) and typically associated with physical infrastructure such as roads[9–11,25]. Linear scaling is often driven by individual needs that are density-independent such as total housing and household electricity consumption[11]. Superlinear scaling (scaling exponents >1) is often the consequence of the densifying social interactions and recent findings typically relate it to intellectual or virtualized features such as patent production or wealth creation[9–11,25].

For waste production, distinct types of waste may follow fundamentally different relationships with urbanization. For example, coal-fired power plants achieve efficiencies with scale in terms of $CO_2$/kWh[28], and thus if larger cities employ larger power plants we might expect sublinear scaling with city size. In contrast, increased social connections and intellectual activity could increase the output of certain types of waste, but for which ones (and why) remains unclear.

Here we synthesize three major forms of human waste in more than 1000 cities across 171 countries. The cities span populations of 50,000 to 24 million people (Fig.1a). We examine Chinese cities as a detailed case study of rapid urbanization where detailed data are available for diverse cities within a single nation.

**Scaling laws globally**

We find universal scaling laws of waste production globally across diverse urban systems spanning all three major forms of waste we considered: wastewater (Fig.1b), municipal solid waste (MSW, Fig.1c), and greenhouse gas emission (GHG, Fig.1d). According to these scaling laws, the overall waste production of a city can be reliably predicted based on city size measured by urban population.

What is unexpected however, is that these scaling laws are functionally distinct across different waste types, likely due to inherently different mechanisms of waste generation. For example, the production of municipal solid waste (Fig.1c) seems to scale linearly with city size ($\beta \sim 1$, $r^2=0.83$, $p<0.001$), that is, doubling the city's population will double its municipal solid waste generation. This linear relationship suggests that municipal solid waste generation is driven by individual needs, independent of the density or size of the city where people reside.

In contrast, both wastewater (Fig.1b) and greenhouse gas emissions (Fig.1d) productions scale nonlinearly with city size ($\beta \neq 1$), that is, doubling the city size will *not* double its generation of waste water or greenhouse gasses. Perhaps most intriguingly, the production of wastewater seems to scale super-linearly with city size ($\beta \sim 1.15$, $r^2=0.79$, $p<0.001$) such that doubling population will more than double waste production. The theory of urban scaling suggests that superlinear scaling in cities is driven by the increased rates of interpersonal interactions as cities densify with increasing population size[29]. This type of scaling leads to greater rates of economic activity such as the creation of wealth[10,30–33] and consequently a greater rate of wastewater generation, because wealth creation (especially in the industrial sector) is strongly coupled with water consumption.

In contrast to wastewater production, greenhouse gas emissions, such as $CO_2$, scale sublinearly with city size ($\beta \sim 0.85$, $r^2=0.47$, $p<0.001$), pointing to a relative economy of scale for the processes that generate greenhouse gasses[34,35]. For example, the energy efficiency of the urban transportation system increases with city size and population density due to the increasing use of public transportation[37]. Note that our finding holds regardless of whether we use only emissions generated within the city boundary (scope-1 emissions), or emissions generated outside of city boundary (e.g., imported grid-power; scope-2 emissions) (Methods; Extended Data Fig.2).

That various types of waste are characterized by significant scaling relationships is important and implies, on average, that waste generation is determined by a common set of organizing principles related to city size globally. While the central relationship accounts for most of the variation, it is also useful to consider the cities that deviate significantly from the power law. Here deviations are most productively considered in terms of the normalized distance from the scaling relationship (e.g. ref [26]). For example, in China there is a gradient in both wealth and aridity as one moves inland from the coast, and one should expect such gradients to alter water use and wastewater production. Indeed Fig.1b shows that Dongguan (high precipitation and high per capita GDP) uses more water than expected and Tianshui (low precipitation and low per capita GDP) uses less water than expected. Similarly, Seattle and Malawi are very similar in population but differ by a factor of roughly 40 in per capita GDP, leading to roughly an order of magnitude difference in municipal solid waste production (Fig.1c). In the next section we evaluate such deviations and unify the variation under a common framework that combines wealth and the natural environment.

**<u>Natural constraints and deviations</u>**
A central idea in evolution and ecology is that local species are adapted to local environments[38]. Traits found in an environment will typically be relatively well-matched to that environment, such that environments with scarce water will have species that use water more efficiently. This idea extends to life-styles and strategies of hunter-gatherer groups and early human settlements[39–41]. For cities, then, to what extent are patterns of waste production influenced by local environmental constraints, or do increasingly global supply chains decouple local constraints from local activities?

We tested this idea by examining the residuals of scaling relationships, as a way to define relative over- and under-performance after accounting for the inherent economies or diseconomies of scale. Consistent with intuition, the amount of rainfall cities receive helps to explain the deviations of wastewater production away from the scaling law: cities with higher rainfall produce more wastewater than expected for their size and cities with lower rainfall use water more efficiently. In addition to this local constraint (*i.e.*, natural precipitation), the production of wastewater is dictated by socio-economic factors such as per-capita GDP. Richer cities generate more wastewater than expected for their size and poorer cities generate less than expected. The combined result is a planar function (Fig.2a) that can explain the city-specific deviations from the wastewater scaling law in Fig.1b.

For the production of municipal solid waste and greenhouse gasses the city-specific deviations from scaling laws are not explained by natural constraints such as rainfall or temperature. Instead, we find that these deviations are explained primarily by per capita wealth (Fig.2b,c). In both cases, increasing wealth will lead to a positive deviation away from the waste scaling law, and thus create cities that are more wasteful. This finding suggests that certain types of waste are emergent phenomena of the internal characteristics of cities rather than being constrained by the natural attributes of the immediate surroundings.

We next examine cities with relatively high residuals (hereafter *wasteful* cities) and low residuals (more *efficient* cities). A detailed knowledge of these cities might inform policy making and future planning. Consistent with our residual analysis, we found that for wastewater generation, more *wasteful* cities on average receive 34% more annual rainfall than relatively *efficient* cities (1030 mm *vs.* 767 mm, p < 0.001; Extended Data Fig.4a). We also found that, for wastewater generation, these *wasteful* cities possess more than double the per capita GDP of these *efficient* cities ( $27164 *vs*. $13312, p < 0.001; Extended Data Fig.4b). For the production of municipal solid waste and greenhouse gasses, we found no significant differences in any of the environmental variables we analyzed, despite strong (3-4 fold) differences in per capita GDP between *wasteful* and *efficient* cities (Extended Data Fig.4b and c).

Together, our results suggest that the waste cities produce can be understood in terms of both the city size (Fig.1) and with the influences of financial and natural resources (Fig.2, Extended Data Fig.4). These results are important for forecasting waste production into the future, as we discuss next.

### Shifts in time

Another important consideration for cities is how waste generation is changing in time (Fig.3). This is especially relevant as nations continue to increase total wealth, urbanize, and implement new technologies that may either lower or increase consumption and waste generation. From a scaling perspective, this can be represented as a change in scaling exponents through time[42](Fig.3a-c). For example, an exponent rising in time indicates that larger cities are increasing waste production faster than smaller cities. We find that wastewater generation (Fig.3a,d) maintains a consistent superlinear scaling. This diseconomy of scale seems to have stabilized, implying that continuous urbanization is expected to radically increase the amount of urban wastewater produced. Greenhouse gas emission (Fig.3c,f) maintains a consistent sublinear scaling in time. Despite the uncertainty associated with a smaller number of cities in the temporal dataset (Fig.3c, Extended Data Table1), cities maintain a relative economy of scale over time for greenhouse gas emissions, consistent with our global analysis in Fig.1d. In contrast, municipal solid waste shifts from strongly sublinear scaling to linear scaling (Fig.3b). In China, where we have temporal data, over time the economy of scale in solid waste production is lost by cities. On the upside, cities of various sizes seem to be equilibrating to near linear scaling in time, which implies that urbanization matters less since people in cities of all sizes produce comparable amounts of solid waste per capita. On the downside, the economies of scale that have been historically achieved for urban solid waste production no longer exist. This trend suggests that cities are drifting away from a future that would realize considerable solid waste reduction with increasing city size.

### Breaking the vicious circle of waste production

Our work provides a framework for understanding current and future human waste production. If cities simply change in population size they should follow the current scaling relationship defining the system they are a part of. If the exponents are sublinear overall, waste per capita should decrease, although total waste will increase given the increasing total population. If the exponents are superlinear, waste will dramatically increase due to both a larger population and larger per capita production. This perspective highlights the startling challenges associated with superlinear products like wastewater which will increase dramatically as cities increase in size. Our results provide a very moderate amount of hope for waste products with economies of scale, such as greenhouse gasses, where per-capita generation is decreasing with city size. However, for net zero emission goals to be achieved all cities need to adjust the overall magnitude of emissions in combination with an economy of scale. Decreasing both the exponent and overall magnitude of urban scaling relationships is required.

Ongoing urbanization in India, sub-saharan African, and other areas of the developing world should increase both city sizes and personal wealth. In particular, the world is expected to increase in population by 2 billion people in the next 30 years (WPP2019, UN), a 26% increase of population. In contrast, urban population is projected to increase by ~60% (WUP2018, UN). Population alone will mean more waste production even for products that grow sublinearly, but for superlinear products population growth combined with increasing urbanization and increasing wealth implies rapid growth in per capita waste (Fig.4c). More concretely, if a current city of 1 million people doubles in size, then wastewater production increases by 122%, and for this city this is equivalent to a shift of 100 million to 221 million tonnes per year.

Fig.4a illustrates the types of trajectories that cities can take and how these will change waste production. Here increases in GDP or population will increase waste production, and the combined increase will change waste production most rapidly which is the scenario most likely to occur globally. Indeed, such trends are observed as it is the case that cities are mostly increasing in size in time and that municipal solid waste production is going up in various size categories of cities (Fig.4b,c).

Increases in GDP often improve quality of life for individuals in cities, so we need ways to decouple the link between increasing economic prosperity and per-capita waste production [43,44]. Perhaps we can look to San Francisco or Japanese cities as examples of high GDP and low waste cities. For roughly the same per capita GDP, Japan generates ⅓ of municipal solid waste per person compared to the United States, and San Francisco generates less per-capita municipal solid waste than any other major city in the US[1]. The structural features, cultural dynamics, and policies that allow these cities to reduce waste need to be more systematically understood in connection with scaling and deployed to most global cities.

**The hidden half of human productive economies**
Most economic and social theories of human civilization, including urban scaling theory, focus on the concepts of production, growth, innovation and the forces that ultimately limit these productive processes. In contrast, the other half of the equation, the byproducts (i.e., waste consequences) of these productive processes have largely been overlooked. Our new work fills this knowledge gap, illuminating the scaling of waste production in urban systems.

This perspective connects to classic ecology and the history of life where the accumulation of waste is seen as a fundamental limit for all species[45,46]. Complete theories of economies must implement similar considerations for human society. Moving forward, a complete science of waste would couple byproducts to products, and an ultimate science of the economy would consider the feedbacks between production and waste including eventual hindrances produced by waste. Notably, this is already considered by many efforts which seek to price the eventual economic costs of global warming due to anthropogenic $CO_2$ emissions[47]. These considerations should be expanded to all types of waste through the lens of urban systems and how waste scales with city size as we have presented here.

Quantifying numerous dimensions of waste will be important for a complete theory of production and waste generation. For example, our quantitative framework should be extended to all types of waste (e.g., medical waste) in order to understand the complete dynamics. According to our country-level analysis, for each billion-ton of municipal solid waste generated, a county on average generates 2 billion tons of construction waste, 600 million tons of agricultural waste, 300 million tons of industrial waste, 100 million tons of hazardous waste, 10 million tons of electronic waste, and 8 million tons of medical waste (Fig.5). These wastes heavily burden our society and natural ecosystems by their sheer volume, but more importantly, threaten public health and safety through polluting air and drinking water[48], clogging the sewage system and creating flooding[49], polluting soil and consequently the entire food web through cascading interactions[50]. Our findings raise the urgent need for a holistic approach to account for all these waste types, and points to lack of city-level waste generation data much needed for future analysis. In addition, future analyses of economic performance must incorporate the costs of waste, for example via natural capital accounting[51], in connection with normalizations that consider city size and local natural resources.

Our work shows the need for a general theory that incorporates the integrated science of economic production, waste generation, and, eventually, the full cycle of material flow. Realizing this goal should enhance humanity's pursuit of a more sustainable future.

## Methods
### Scaling theory
Urban scaling theory relates city size to a given feature of interest through power law relationships of the form $Y = Y_0 N^{\beta}$ [9,11]. The power of this perspective is that often the exponent $\beta$ is not equal to one, demonstrating nonlinearities that contradict simple per-capita perspectives along with an economy or diseconomy of scale, and typically the exponent reveals something about underlying mechanisms. For example, superlinear exponents are the consequence of densifying social interactions with urban environments and are typically related to intellectual or virtualized features such as knowledge creation or wealth production[10,11,30,32,33]. At the most basic level there are more people per unit area in larger cities, such that interactions are easier and the total interaction rate between individuals increases with city size [29]. Sublinear exponents are typically related to infrastructural economies of scale such as the total area of roads or total number of gas stations[10,52,53]. A powerful aspect of this theory is that it allows us to transcend the individual and consider the nontrivial effects of larger collections of humans[9]. In addition, the scaling relationships often provide a nontrivial baseline against which to measure deviations and variance[26]. These deviations reveal the character of individual cities and elucidate higher-order mechanisms beyond the scale of a city.

### Data compiling
We compiled 5 city-level waste datasets from 4 different sources (Extended Data Table.1) that encompasses three major waste types: wastewater, municipal solid waste (MSW, colloquially known as "city trash"), and greenhouse gas emissions (GHG, $CO_2$ equivalent). Each dataset contains urban population data. We used a minimum threshold of 50,000 urban residents for defining a city[54]. The definition of a city/urban system is consistent within each scaling analysis of each dataset. These datasets, detailed below, together enable us to do cross-sectional analysis (Fig.1,2) and temporal analysis (Fig.3,4) for these three waste types.

*MoHURD_wastewater* and *MoHURD_MSW*: we acquired a centralized database curated by the Ministry of Housing and Urban Rural Development (MoURD, China; Extended Data Fig.1b). This database contains urban population census and city-level waste production data from 2002 to 2020. From this centralized database we extracted and homogenized wastewater and MSW datasets that covered city-specific waste production time-series from 2006-2020.

*Worldbank_city*: we accessed city-level municipal solid waste data (Extended Data Fig.1a) from the World Bank (https://datacatalog.worldbank.org/search/dataset/0039597).

*Nangini2019*: We acquired this global dataset on city-level greenhouse gas emissions[55] (Extended Data Fig.1c). This dataset contains only cross-sectional data, with no temporal series. The metric of greenhouse gas emissions we used is territory-based, including transport, industrial, and local power plant emissions within the city boundary (commonly referred as scope-1 emissions[56]). A consumption-based emission metric will further include emissions embedded in traded goods as well as grid-supplied energy consumption produced by power plants outside the city boundary (commonly referred to as scope-2 emissions). At the spatial resolution of city scale, data on scope-2 GHG emissions is scarce and less reliable as it is much harder to derive and often involves a range of assumptions[55]. We examined our finding of sublinear scaling of GHG using the total emissions data (scope-1 + scope-2) included in this dataset. Despite a much smaller sample size (n=130) and only covering cities from developed countries (USA, Canada, Australia, New Zealand and Europe), the scaling relationship of total emissions is consistent with our main findings (Extended Data. Fig.2)

*C40*: We acquired a temporal dataset of greenhouse gas emission from cities of the C40 Cities Climate Leadership Group program (C40; Extended Data Fig.1d). Consistent with the cross-sectional analysis in Fig.1d, we used scope-1 emissions (*i.e.*, territorial) for the temporal analysis (Fig.3). The number of cities reporting emissions data varied over time: less than 3 cities during 1990-2004, 4~7 cities during 2005-2011, 17~42 cities during 2012-2019, and 5 cities for 2020. Scaling analysis is constrained to 2012-2019 due to paucity of data in other years (Fig.3, Extended Data Fig.5).

For the residuals analysis of scaling laws (Fig.2), we compiled city-level GDP and climatic variables for each city in our dataset compiled above. We acquired data on city-level GDP from two sources: (1) Oxford economies dataset (proprietary) that spans 902 cities worldwide from 2000 to 2020, and (2) a GDP dataset from Zhou's lab that features 292 Chinese cities spanning 1988 to 2018. All GDP data were converted by purchasing power parity (GDP_ppp) before being used in our analyses. Combined, we were able to integrate these city-level GDP data with our existing waste data. For environmental factors, we acquired the following abiotic conditions for each city based on their geo-coordinates: (1) Temperature, (2) annual temperature range, (3) Annual precipitation, (4) Dry season precipitation, (5) Precipitation seasonality and (6) Aridity. We derived elevation from the Google Elevation application programming interface. All climatic variables were derived from the 30-year average (1970-2000) at 1km resolution from WorldClim Version 2[57].

We also compiled 3 country-level waste datasets from 3 different sources (Extended Data Table.1) to generate country-level accounting of waste generation (Fig.5): (i) *Jones2021*, a recent global analysis of country-level wastewater generation[58]; (ii) *Ritchie2020*, a global analysis of country-level greenhouse emissions[59]; and (iii) *Worldbank_country*, a country-level dataset curated by the World Bank that covers 7 waste types: MSW, construction, agricultural, industrial, hazardous, electronic, and medical waste. The greenhouse dataset (*Ritchie2020*) contains time-series whereas *Jones2021* and *Worldbank_country* are cross-sectional. To facilitate comparison among datasets (Fig.5), we randomly sampled the *Ritchie2020* dataset between 2009 to 2018, such that the resulting country-level data has an ensemble of years that are similar to the time distribution of other datasets. The combined dataset is summarized in Fig.5, Extended Data Table.2, and available on figshare (Data availability).

**Statistical analyses**
We used linear regression to analyze scaling relationships in Fig.1,3. Both waste production rate (tonnes/year) and population size are log10 transformed, consistent with logarithmic scale representation in each figure panel. Residuals (in log10 units) of the scaling relationships derived in Fig.1 are then fed into Fig.2 as the dependent variables. Residuals (for each city) that have no matching environmental or socio-economic data are dropped from the analysis. For the planar function of wastewater production in Fig.2a, we tested no interaction effect between per capita GDP and other environmental variables. We analyzed the magnitude difference across 9 waste types (Fig.5) using linear regression (*lm*; R package "stats") followed by pairwise contrast analysis (*emmeans*; R package "emmeans"). We log10-transformed country-level waste generation rate, consistent with the logarithmic scale of the y axis in Fig.5. For testing feature differences between *wasteful* vs. *efficient* cities (Extended Data Fig.4), per capita GDP was first log10-transformed to conform with normality. We tested equal variance using the F-test before applying t-test: Welch's t-test in the case of unequal variance (Extended Data Fig.4a,b), and Student's t-test in the case of equal variance (Extended Data Fig.4c,d). All analyses were conducted in R (v4.2.0).

**Code availability**
R scripts are available from the corresponding author upon request (mingzhen.lu@santafe.edu).

**Data availability**
All data will be deposited at figshare upon publication of this paper.

**Author contribution**
M.L. developed the overall conceptual framework and analyses with input from C.K. and R.B.J.  M.L compiled and analyzed the data, with input from C.K., C.W., and R.B.J..  Z.C. provided a GDP dataset for Chinese cities. M.L and C.K. wrote the first draft and all authors contributed to revisions. The authors declare no competing financial interests.

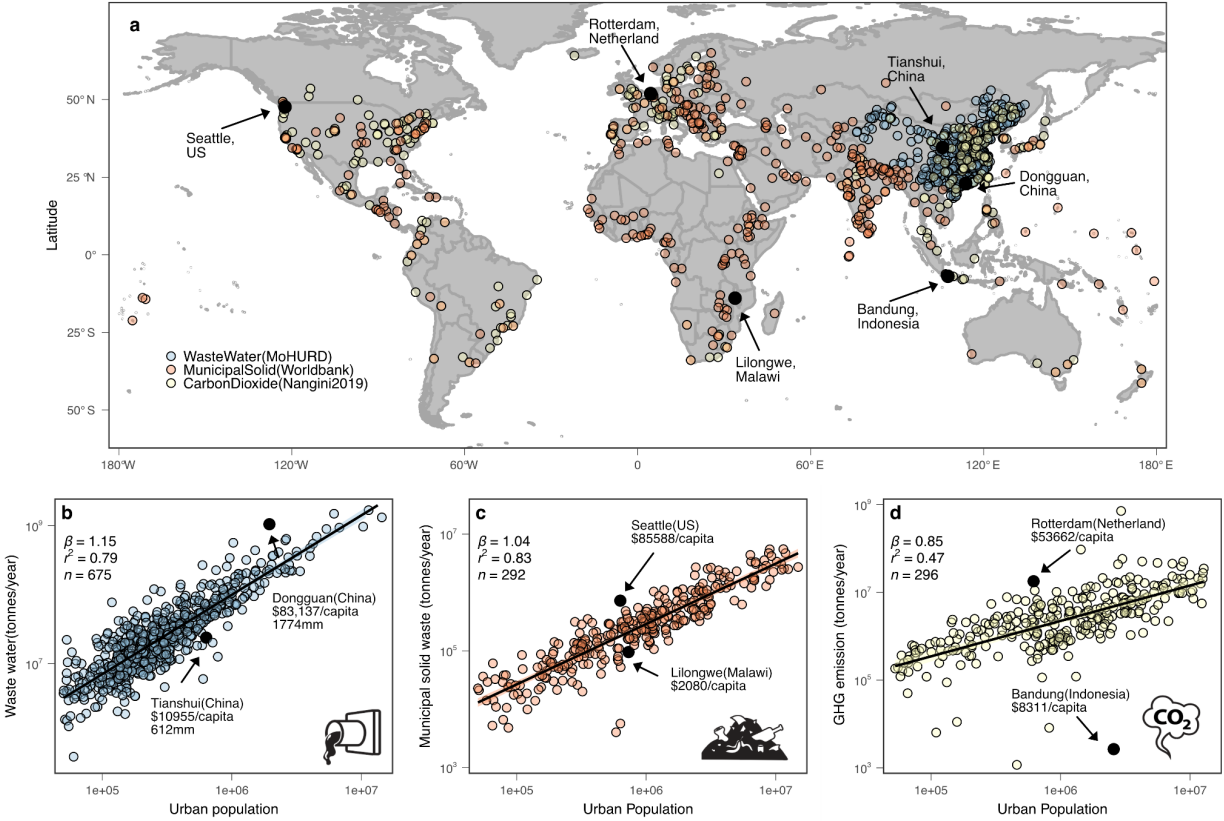

**Figure 1|Scaling law of waste production across cities worldwide: the super-linear, the linear, and the sub-linear.** (**a**) Geolocation of cities from 3 distinct datasets of waste: (1) City-level dataset of waste water from the Ministry of Housing and Urban Rural Development (MoHURD), China (blue point, number of cities n = 712); (2) City-level municipal solid waste (MSW) data from Worldbank (orange point, number of cities n = 363, ref dataset); and (3) city-level greenhouse gas (GHG) emissions compiled by Nangini et al. (2019) (yellon point, n = 343). (**b**) The generation of waste water increases non-linearly with increasing city size. A cross-sectional dataset composed of single-year snapshots suggests a super-linear relationship between size of cities and their wastewater production ($\beta = 1.15 \pm 0.04$). We highlight two example cities that stand out with large deviation from the scaling law (black dots). Dongguan, an industrial city of Southern China which features high personal wealth and high annual precipitation, generates disportionately more waste water given its size. In contrast, the northwestern city of Tianshui, which features much lower personal wealth and rainfall, generates much less waste water given its size. (**c**) The municipal solid waste production scale linearly with city size ($\beta = 1.04 \pm 0.05$, n = 292). The linear scaling indicates lack of interaction among individuals producing municipal waste, consistent with the theoretical prediction of the "individual/household" process[29]. We highlighted Seattle (US) and Lilongwe (Malawi) that deviates away from the general scaling relationship. The much richer Seattle produces 8 times more municipal waste compared with Lilongwe, even though it has a smaller population size. (**d**) The emission of greenhouse gases (GHG) displays sublinear scaling across cities worldwide ($\beta = 0.85 \pm 0.1$, *n*=296). We highlight Rotterdam (Netherland) and Bandung (Indonesia) that deviates from the general scaling relationship, with Rotterdam producing disproportionally more GHG. Error bands in (b)-(d) represent the confidence interval of each scaling relationship.

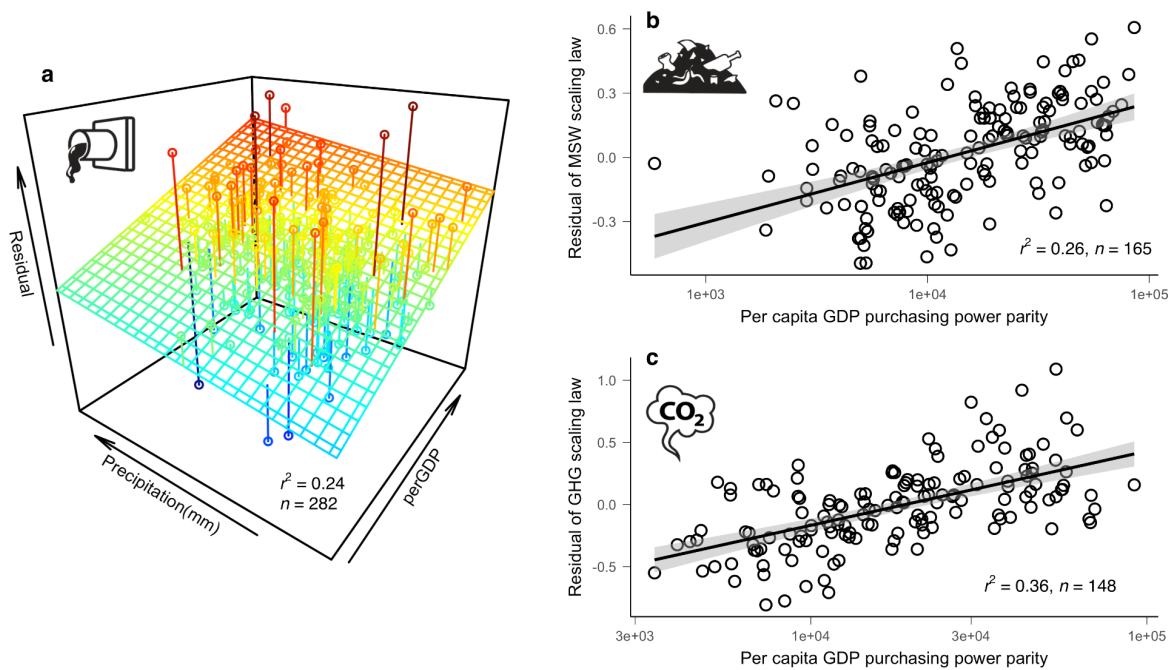

**Figure 2| Deviation from scaling laws explained by environmental and economic factors. (a)** The residuals of the wastewater scaling relationship (Wastewater= 1.1 * UrbanPopu^1.15 in Fig.1b) can be explained by the combination of annual precipitation and per capita GDP (purchasing power parity) using a planar function (Residual =-1.6+ 0.35*log10(perGDP) + 1.003e-4*Precip; $r^2$ = 0.24, n = 282). The richer the city and the more rainfall it receives, the city will deviate from the scaling law more positively. **(b)** The residual of municipal solid waste scaling law can be explained by per capita GDP (purchasing power parity) in a non-linear relationship (Residual = -1.14+0.28*log10(perGDP); $r^2$ = 0.26, n = 165). **(c)** The residual of GHG scaling law can be explained by per capita GDP (purchasing power parity) in a non-linear relationship (Residual = -2.56+0.6log10(perGDP); $r^2$ = 0.36, n = 148). Note that we removed 3 outliers from this regression. Keeping the three outliers would not alter the slope of the regression (b = 0.6), but would shift downwards the intercept (a=-2.65). See Extended Data Figure 2 for a version of the regression in which we retained the outliers. Error bands in (b) and (c) represent the confidence interval of each regression fit. Note that the x-axis is in logarithmic scale while the residual value on the y-axis is in log10 unit.

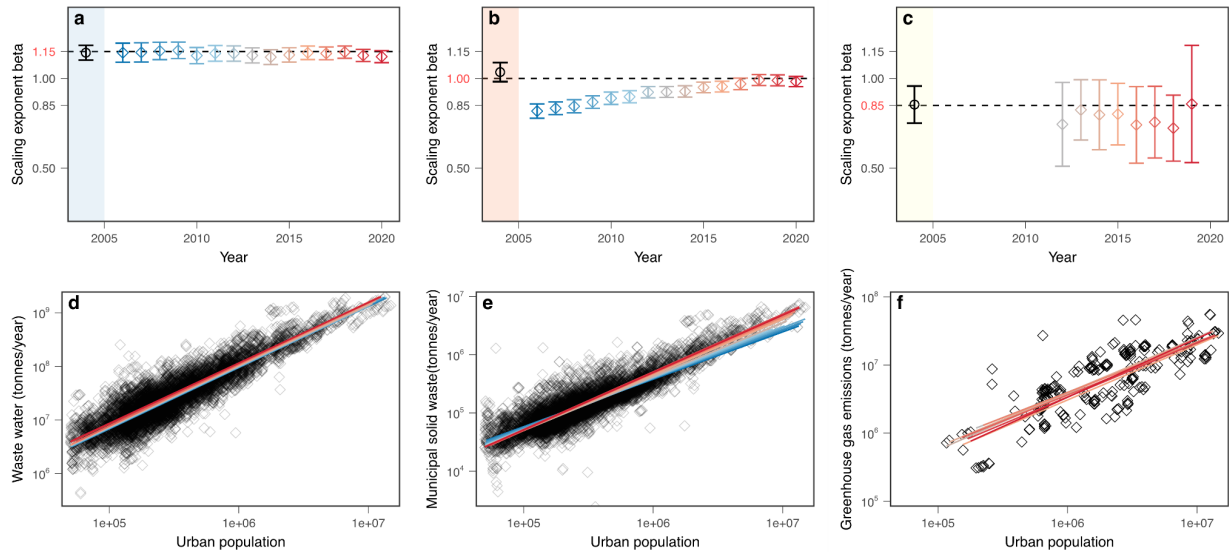

**Figure 3| Temporal evolution of scaling relationship.** (**a**) Scaling exponent of wastewater production hovers around 1.15 over the last two decades for 600+ Chinese cities (MoHURD dataset). Black dot represents the scaling slope from a cross-sectional scaling relationship as in Fig.1**b**, whereas the color gradient from blue to red dots indicate scaling slope of each year from panel (**d**). (**b**) The scaling exponent of municipal solid waste started as sublinear but over time gradually approached linear (MoHURD dataset), approaching the global average (black dot) as in Fig.1**c**. (**c**) A temporal analysis of the C40 dataset displays higher variation, but the estimated exponents consistently hover around 0.85. Due to the small size of the temporal dataset, uncertainty of the scaling exponents is large, yet the central tendency corroborates with the sublinear scaling of global analyses in Fig.1**d** (black circle). Scaling analyses in panel (**c**) were restricted to years when data were available from more than 15 cities (i.e., 2012-2019). Extended Data Fig.5 displays raw data from 2005-2020.

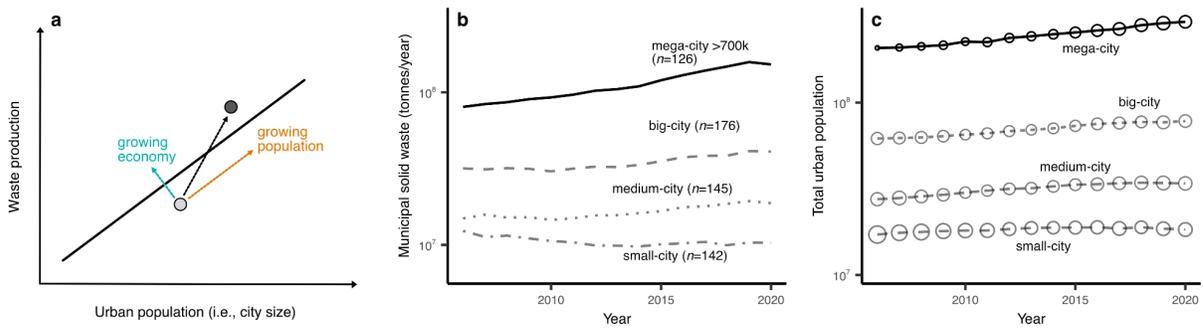

**Figure 4| Understanding the forces that drive waste generation across cities and through time.** (**a**) A city that is currently situated at the lower left of the worldwide waste scaling line can change in two orthogonal directions: (1) change of urban population, shrinking or growing, parallel to the scaling line; and (2) change of efficiency orthogonal to the scaling line. In the example we give, the underdeveloped city denoted in gray dot increased in population and increased their per-capita waste generation (i.e., decreasing efficiency, driven by economic growth), and consequently shifted to the upper right of the scaling line. (**b**) When summing up the total municipal solid waste generation of Chinese cities according to their size classes, our analyses identify that the biggest cities feature the fastest growth rate of waste production, while the smallest cities feature the slowest growth rate of waste production. City size classification is based on urban population of year 2020: mega-city (urban population x>700k, number of cities belonging to this category n = 126), big-city (300k<x<700k, n=176), medium-city (180k<x<300k, n = 145), and small-city (50k<x<180k, n = 142). (**c**) The rapid growth of waste generation in large cities is driven both by China's rapid urbanization in the past 15 years, and a steady increase of per-capita waste generation rate in the largest cities (size of circle). Note that small cities actually decreased in per-capita waste, but the total population in small cities is magnitude smaller than that of bigger cities. As a result, we observe an increasing scaling slope in Fig.3b.

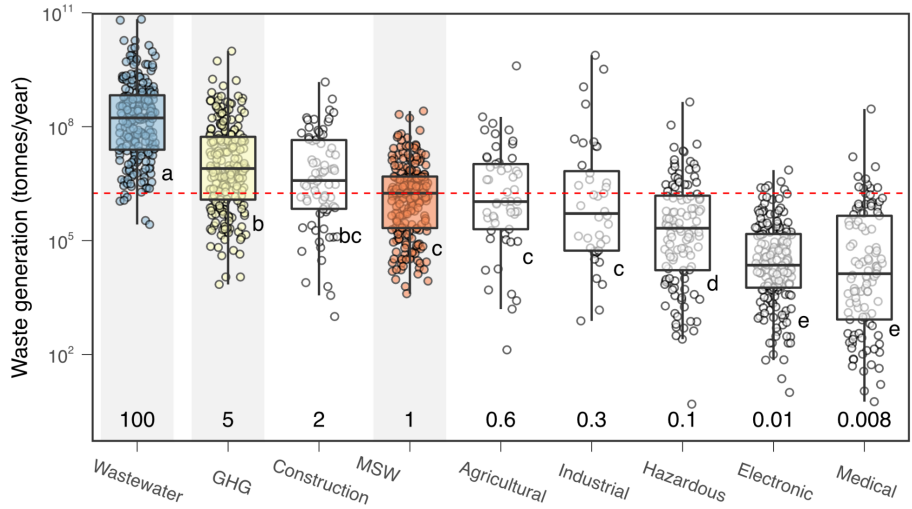

**Figure 5| Waste generation and global accounting of country-level waste generation.** Country-level waste generation is ranked according to their median value across 9 major types of waste: waste water, greenhouse gasses (GHG)carbon dioxide, construction waste, municipal solid waste (MSW), agricultural waste, industrial waste, hazardous waste, electronic waste, and medical waste. Each point represents a total for a country. The bottom row of numbers represents the ratio of country-level medians worldwide relative to municipal solid waste. The median value of country-level municipal solid waste production is ~1.8 million tonnes per year indicated by a red dashed line. Compact letters indicate significant pairwise differences among waste types (α = 0.05). Waste water, carbon dioxide, and municipal solid waste (shaded gray areas) are the focus of our city-level analyses, and are highlighted with color consistent with other figures. We used white fill to indicate waste types that we do not have city-level data for.

# Extended Data Tables and Figures

**Extended Data Table 1: Summary table of datasets analyzed**

| Dataset | Waste type | Temporal | Resolution | Scope | Related |
|---|---|---|---|---|---|
| **(a) City-level** | | | | | |
| MoHURD_Wastewater[a] | Wastewater | Time-series | City | China | Fig.1, Fig.2, Fig.3; ED Fig.1,4 |
| Worldbank_City[b] | MSW | Cross-sectional | City | Global | Fig.1, Fig.2; ED Fig.1,4 |
| Nangini2019[c] | GHG | Cross-sectional | City | Global | Fig.1, Fig.2; ED Fig.1-4 |
| MoHURD_MSW[a] | MSW | Time-series | City | China | Fig.3, Fig.4; ED Fig.1 |
| C40[d] | GHG | Time-series | City | Global | Fig.3; ED Fig.1,5 |
| **(b) Country-level** | | | | | |
| Jones2021[e] | Wastewater | Cross-sectional | Country | Global | Fig.5 |
| Ritchie2020[f] | GHG | Time-series | Country | Global | Fig.5 |
| Worldbank_Country[b] | MSW, Construction, Agricultural, Industrial, Hazardous, Electric, Medical | Cross-sectional | Country | Global | Fig.5 |

"MSW" stands for municipal solid waste, "GHG" stands for greenhouse gasses. [a]Centralized urban datasets curated by the Ministry of Housing and Urban Rural Development (MoURD, China). The wastewater and municipal solid waste datasets are extracted, homogenized. [b]"What a waste" datasets created by the World Bank[5]. The city-level dataset contains only data on municipal solid waste, while the country-level dataset covers a broader range of waste types. [c]A published dataset on greenhouse gas emissions from Nangini et al. (2019)[55]. [d] A temporal dataset of greenhouse gas emissions from cities of the C40 Cities Climate Leadership Group program (C40). [e]A recent global analysis of country-level wastewater generation by Jones et al. (2021)[58]. [f]A global analysis of country-level greenhouse emissions[59].

**Extended Data Table 2: Waste generation statistics across countries worldwide.**

| WasteType | N | Median | Min | Quantile_25 | Quantile_75 | Max | Mean | WasteRatio |
|---|---|---|---|---|---|---|---|---|
| Wastewater | 215 | 170119799 | 265782 | 24760813 | 672945942 | 6.762E+10 | 1671649270 | 100 |
| GHG | 215 | 8163000 | 4000 | 1140000 | 53356000 | 9920459000 | 157278470 | 5 |
| Construction | 74 | 3820575 | 1000 | 687402 | 44474597 | 1.50E+09 | 57779234 | 2 |
| MSW | 215 | 1768977 | 3989 | 213879 | 4865910 | 2.58E+08 | 8582534 | 1 |
| Agricultural | 53 | 1063964 | 133 | 199453 | 10365788 | 4.00E+09 | 90015688 | 0.6 |
| Industrial | 40 | 520396 | 766 | 54312 | 6900000 | 7.60E+09 | 315096014 | 0.3 |
| Hazardous | 118 | 212020 | 5 | 16599 | 1524000 | 4.48E+08 | 6934517 | 0.1 |
| Electronic | 182 | 22500 | 10 | 5750 | 149250 | 7211000 | 242295 | 0.01 |
| Medical | 105 | 13408 | 6 | 834 | 450000 | 2.92E+08 | 3399483 | 0.008 |

N indicates numbers of countries in the dataset. All major stats (from Median to Mean) are in unit tonnes/year. "GHG" stands for greenhouse gasses, "MSW" stands for municipal solid waste. Waste ratio is calculated by dividing Median with the median value of municipal solid waste.

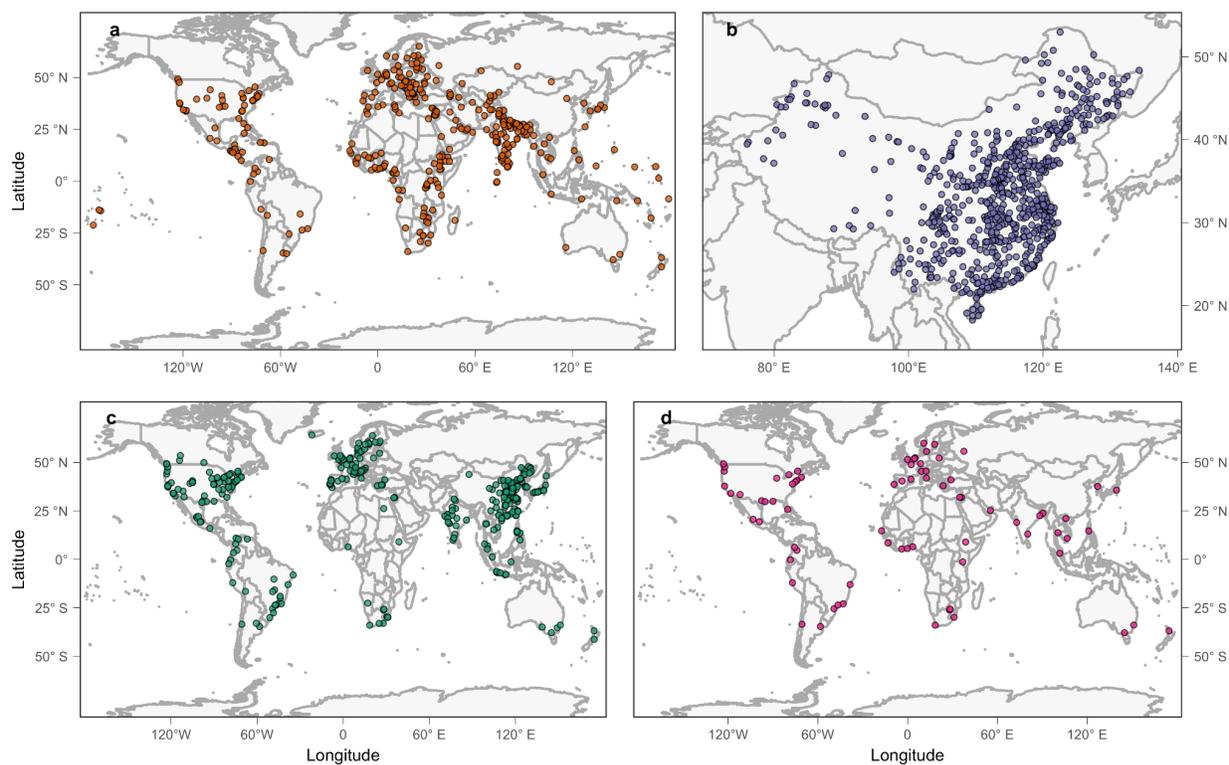

**Extended Data Figure 1**: Geolocation of cities in our datasets from four different sources: world bank waste dataset(**a**), MoURD dataset for China (**b**), GHG emission dataset from Nangini et al. (2019) (**c**) temporal GHG emission dataset from C40 (**d**). More information can be found in Methods, section "Data compiling".

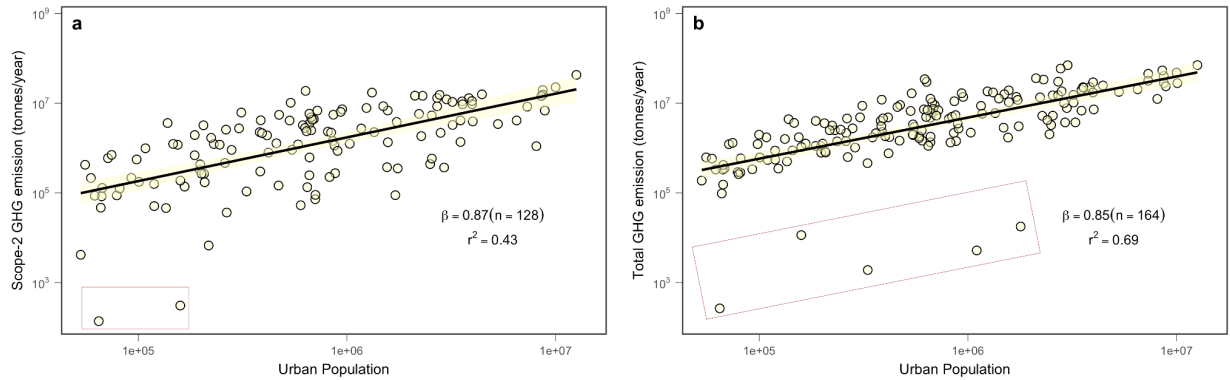

**Extended Data Figure 2 | Sublinear scaling of GHG across different metrics of emissions.** (**a**) Consistent with our main analysis in Fig.1d (emissions within the city boundary), scope-2 greenhouse gas emissions (e.g., grid-power from outside the city boundary) also displays sublinear scaling ($\beta$=0.87, $r^2$=0.43). Note that this is a much smaller dataset (n=128) that only includes cities from developed countries due to limitation of data availability (Methods). (**b**) Total greenhouse gas emissions (sum of scope-1 and scope-2) also scale sublinearly with urban population ($\beta$=0.85, $r^2$=0.69). Similar to panel (a), this dataset is a restricted subset of what is presented in main text Fig.1d. Outliers are presented here (rectangle box) but not included in the statistical test.

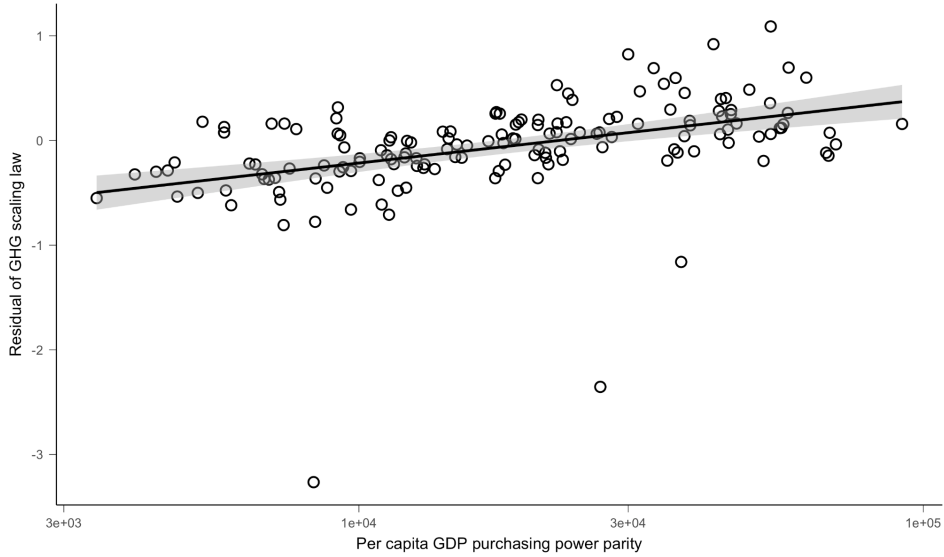

**Extended Data Figure 3| The city-specific deviation from GHG scaling law can be explained by per capita GDP.** The nonlinear relationship takes the form: Residual = -2.65+0.6log$_{10}$(perGDP). Three cities feature exceedingly low log residual, which lead to the low goodness of fit ($r^2$ = 0.18, n = 151). Removing these 3 outliers would greatly improve the fit, but does not impact the slope of the fit (Fig.2c). Error band represents the confidence interval of the regression fit. Note that the x-axis is in logarithmic scale while the residual value on the y-axis is in log10 unit.

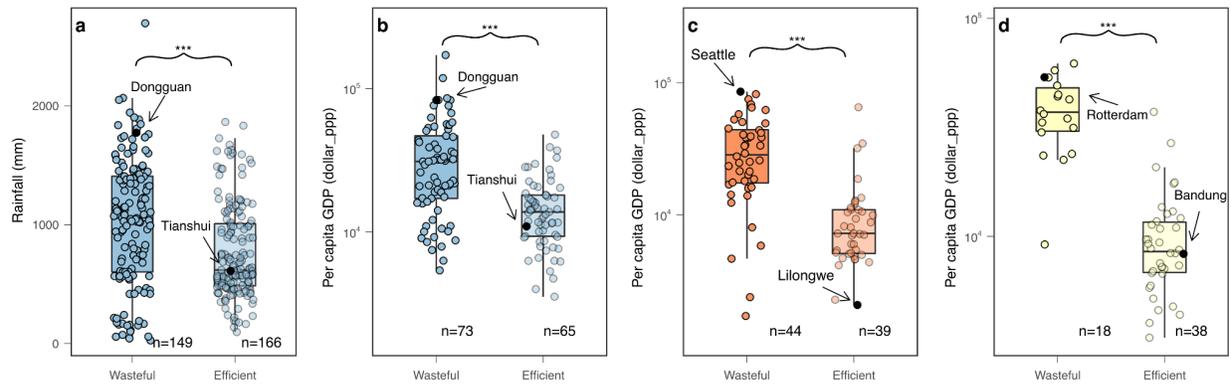

**Extended Data Figure 4| Comparison of city properties between the *wasteful* and *efficient* cities.** (**a**) Cities that produce more waste water on average feature higher annual rainfall (1030 mm *vs.* 767 mm, Welch's t-test, p < 0.001). (**b**) Cities that generate more waste water on average feature higher per capita GDP ($27164 *vs.* $13312, p < 0.001). (**c**) Cities that generate more municipal solid waste feature higher per capita GDP ($25246 *vs.* $7990, p < 0.001). (**d**) Cities that produce more GHG feature higher per capita GDP ($35308 *vs.* $8955, p < 0.001). For all panels, each filled circle represents a single city, and the color fill denotes the type of waste (wastewater in blue, MSW in red, and GHG in yellow).

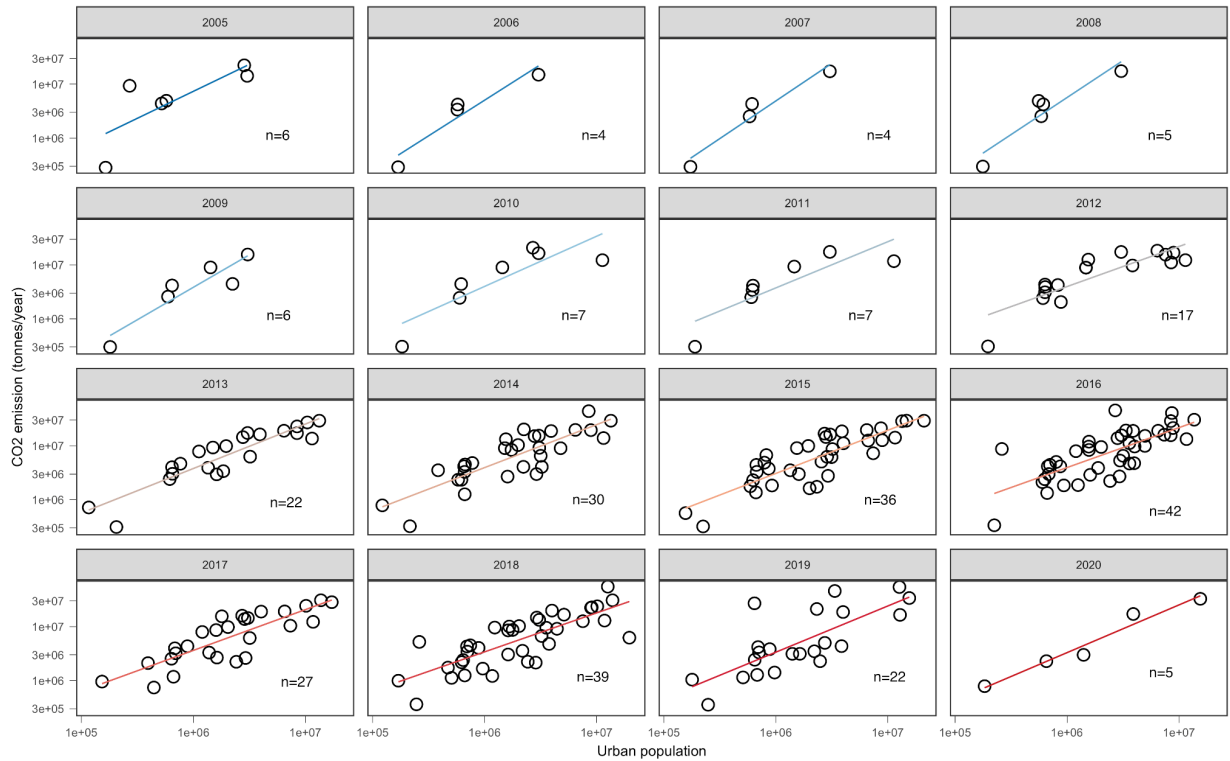

**Extended Data Figure 5| Greenhouse gas emissions of C40 cities broken into different years.** The C40 member cities that are reporting GHG emission data have been steadily increasing from 2005 onwards (less than 3 cities during 1990-2004). We selected a sample size cutoff of 15 cities as a threshold below which scaling analysis is deemed invalid. Using that threshold, we excluded the data year 2005-2012 and 2020. Our treatment is justified because a scaling analysis is a system-level analysis of the entire system of cities. Consequently, a small sample size would fail to give a representative picture of the whole urban system.

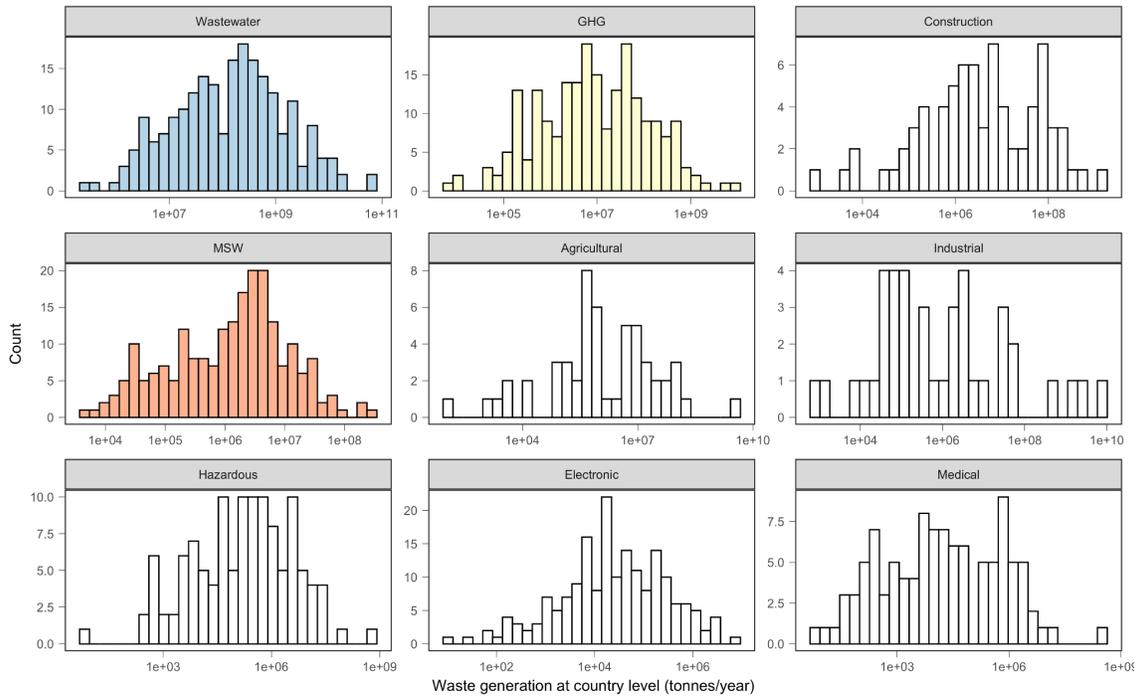

**Extended Data Figure 6| Distribution of country-level waste generation rate (tonnes/year) across major waste types**. The color scheme is consistent with figures in the main text: wastewater in blue, municipal solid waste in orange, and greenhouse gasses in yellow. We used white fill to indicate waste types that we do not have city-level data for.

26